\documentclass[prl,aps,twocolumn,singlespace]{revtex4-1}
\usepackage{graphicx}
\newcommand {\be}{\begin{equation}}
\newcommand {\ee}{\end{equation}}
\newcommand {\bea}{\begin{eqnarray}}
\newcommand {\eea}{\end{eqnarray}}

\begin{document}
\linespread{1.0}
\title{Non-equilibrium local pair potential enhancement}
\author{Ilya Grigorenko}
\affiliation{Department of Physics and Materials Research Institute, Pennsylvania State University, University Park,  PA 16802, USA}
\author{Herschel Rabitz}
\affiliation{Chemistry Department, Princeton University, NJ 08544,
USA}
\date{\today}

\begin{abstract}
We propose a new mechanism for the local pair potential enhancement  
with the help of electromagnetic control fields.
 The mechanism is based on the
creation of non-equilibrium, spatially localized Bogoliubov quasiparticle excitations, which result
in a significant enhancement of the local pair potential and  the local transition  temperature $T_c$.
\end{abstract}

\maketitle
{\setlength{\baselineskip}{1.0\baselineskip}

A properly cooled Fermi atomic gas with an attractive interaction
between atoms can undergo a superfluid transition analogous to the BCS transition in superconductors \cite{stoof}.
This theoretical prediction motivated many experiments  to study  the superfluid properties
of cold Fermi systems \cite{jila,bcs-bec} along with additional theoretical work
 (see, for example, \cite{strinati} and the references herein). Experimental studies have been carried out  on 
the manipulation of 
cold Fermi gases \cite{exp1} based on the control of the critical transition temperature $T_c$. 
Control has been achieved by  application of an external magnetic field \cite{bcs-bec,exp1}, which
 can modify the effective interaction between atoms due to 
the Feshbach resonance \cite{fesh}. The enhanced interatomic interaction may produce a stronger pairing potential 
and a higher $T_c$.

In this work we  propose a new nonequilibrium
mechanism for pair potential enhancement, which is not based on the direct control
of the effective interaction between fermions, but rather relays on
control of the spatial localization of the quasiparticle density in the system.
The local pair potential $\Delta(x_0,t_0)$ at a given position $x_0$ and time $t_0$ is
proportional to the product of the local amplitudes  of the quasi-electron and quasi-hole (Bogoliubov) 
excitations $v_i^*(x_0,t_0)u_i(x_0,t_0)$ \cite{song}. If an external control field is chosen in a such way that it 
drives a significant amount of the
quasiparticles to be localized at a certain moment in the vicinity of  $x_0$, this can  lead to
 an enhancement of the local pair potential $\Delta(x_0,t_0)$.  Therefore, the pair potential can be
enhanced locally in a target volume rather than in the whole system, and this enhancement is
achieved along with  decreasing of the quasparticles density and the local pair potential
elsewhere. This will in effect, result in increase of the 
local effective critical transition temperature $T_c(x_0,t_0)$.
 A simple estimate of the  enhancement of the transition temperature at a given moment $t_0$
can be obtained using the BCS expression for bulk superconductors $T_c(x_0,t_0)\approx \Delta(x_0,t_0)/(1.76 k_B)$.

The suggested  mechanism should be contrasted with the  known
Eliashberg mechanism of the pair potential enhancement
in superconductors \cite{eliashberg_theor}.
Since the time-dependent electromagnetic field may create a non-equilibrium
distribution of quasiparticles in the superconductor, it may lead to
unoccupied states at the Fermi surface at the gap edge that
effectively increases the gap value. However, the Eliashberg mechanism does not result in a {\it spatially localized}
enhancement of the quasiparticle density.
The  Eliashberg mechanism is rather weak effect, resulting in a relative increase of the transition temperature $T_c$ 
by an order of $1\%$ , and it was observed and extensively studied in the cases of
electromagnetic, acoustic or tunneling processes \cite{enhancement_review}.

To illustrate our approach, let us consider the interaction a cold atomic Fermi gas with a time dependent  control
field.  A nonequilibrium state of the system is described utilizing 
time-dependent Bogoliubov-de Gennes (TDBdG) equations for an inhomogeneous
system \cite{song}:
\begin{eqnarray}
\label{tdbdg} i\hbar\frac{\partial u_n({\bf r},t)}{\partial
t}=H u_n({\bf r},t)+\Delta({\bf r},t)v_n({\bf r},t),\\
i\hbar\frac{\partial v_n({\bf r},t)}{\partial t}=-H v_n({\bf r},t)+\Delta^*({\bf r},t)u_n({\bf r},t),\nonumber
\end{eqnarray}
with  $H=H_0+U({\bf r},t)+V({\bf r},t)-\mu$, $H_0$  defined by
\begin{eqnarray}
\label{hamiltonian} H_0=-\frac{\hbar^2}{2 m}\nabla^2+W({\bf r}).
\end{eqnarray}
Here $m$ is electron mass, $\mu$ is the chemical potential, $W({\bf r})$ is an external trapping potential, 
and $V({\bf r},t)$ is
the control field. $U({\bf r},t)$ is the Hartree-like mean field  local
potential, given by
\begin{eqnarray}
\label{u} U({\bf r},t)=-D_0\sum_n [|u_n({\bf r},t)|^2 f_n+|v_n({\bf
r},t)|^2 (1-f_n)].
\end{eqnarray}
 And $\Delta({\bf r},t)$ is the quantity of our interest:
the local pair potential, which is  
\begin{eqnarray}
 \Delta({\bf r},t)=D_0\sum_n u_n({\bf r},t) v^*_n({\bf r},t) (1-2f_n).
\end{eqnarray}
 Here $D_0$ is the effective attraction interaction
coefficient.
For example, for isotropic attractive interactions between neutral atoms in a trap $D_0=4\pi\hbar^2/m a_s$, where $a_s$
is the atomic scattering length. 

To illustrate our approach let us consider a system consisting of a cooled atomic Fermi gas trapped in a potential $W(\bf{r})$.
The trapping potential can be realized, for example, using the induced dipole potential laser trap \cite{ohara}. 
In this case the potential is proportional to the time average of the local laser intensity $I(\bf{r})$. The intensity of 
laser can be modulated, which is equivalent to adding a time-dependent external control potential
$V({\bf r},t)$. For simplicity we assumed that the system is elongated and that the control field $-\nabla V({\bf r},t)$ 
is linearly polarized along the same direction. Thus, the one dimensional description is adequate. For simplicity we 
set the trapping potential $W(x)$ to be
 a square well of the length $L$ with infinite walls. However, the suggested control strategy will work for other types 
of anharmonic 
trapping potentials with non equidistant
 transition frequencies. The system is assumed to be initially in the ground
state. 
We used a mesh with $m=16$ and $m=32$ equidistant discretization points for the spatial variable to perform 
numerical integration of Eq.(\ref{tdbdg}).
 We test our numerical
routine by calculating  
the interaction of a resonant field with  a single particle in an infinite square well potential and compared with the
 analytical solution for the Rabi oscillations between two different states.
We found an excellent agreement between numerical and analytical solutions.

Optimal control of the local pair potential we formulate as a search for a field $V(x,t)$,
 which maximizes the 
average value of the pair potential $<\Delta>$ in a target area $[x_0-\epsilon_0,x_0+\epsilon_0]$ over a given time interval $t_0$:
\begin{equation}\label{delta} <\Delta>=(2 t_0\epsilon_0
)^{-1}\int_{T-t_0}^{T} \int_{x_0-\epsilon_0}^{x_0+\epsilon_0}
 |\Delta(x,t)| dx dt.
\end{equation}

In order to understand how one can control the quantity  Eq.(\ref{delta}), we do the following analysis. 
 The TDBdG equations Eq.(\ref{tdbdg}) can be approximately solved in the absence of the control field ($V(x,t)\equiv0$), 
assuming  constant pair 
$\Delta(x)= \Delta_0$ and Hartree $U(x)= U_0$ potentials. 
The solution in this case is
\begin{eqnarray}\label{anayt}
 u_n(x,t)\approx \bar{u}_n \psi_n(x) e^{-i \frac{E_n t}{\hbar}},
v_n(x,t)\approx \bar{v}_n \psi_n(x) e^{-i \frac{E_n t}{\hbar}},
\end{eqnarray}
 with $\bar{u}^2_n=\frac{1}{2}\left(1+\frac{\epsilon_n-\tilde\mu}{E_n}\right)$,
$\bar{v}^2_n=\frac{1}{2}\left(1-\frac{\epsilon_n-\tilde\mu}{E_n}\right)$,
 where $\tilde\mu=\mu-U_0$.
The eigenenergies are $E_n=\sqrt{(\epsilon_n-\tilde\mu)^2+\Delta_0^2}$,
where $\epsilon_n$ and $\psi_n(x)$ are the eigenenergies and eigenfunctions for the 
stationary Schroedinger equation with the Hamiltonian Eq.(\ref{hamiltonian}).
A control field that maximizes $<\Delta>$ should drive at least some of the amplitudes $u_n, v_n$ from their
initial state $u_n(x,0), v_n(x,0)$
 to a nonequilibrium state at the 
target time    $u_n(x,T), v_n(x,T)$ (assuming $t_0\ll T$), which will have a ``bump'' at, or in a close vicinity of 
$x_0$.

As  initial guess
for the control field $V(x,t)$  we choose one that drives the initial quasiparticle
wavefunctions with the lowest energy $E_1$, $u_1(x,0),v_1(x,0)$, to a nonequilibrium 
 localized  state  $u_1(x,T),v_1(x,T)\propto g(x)$, where  $g(x)=(x-x_0) e^{-(x-x_0)^2/\alpha^2}$, which has two maximums  near 
$x_0=L/2$.
The analytical solution for this
 optimal control field in the dipole approximation can be written using 
the results for localization of a particle in an infinite well potential
 obtained in \cite{squeezing}: $V(x,t)=x f(t)$,
\begin{eqnarray}
\label{field}
f(t)=\sum_{n=2,4,..}^{N}V_n \cos(\omega_n t/\hbar),
\end{eqnarray}
where $\hbar\omega_n=E_n-E_1$, $V_n=\frac{\hbar\pi a_n}{d_n T}$,
$a_n=\int g(x)\psi_n(x) dx$
are the expansion coefficients for the target nonequilibrium state $g(x)$ in the basis $\psi_n(x)$, and 
$d_n=\int \psi_1(x) x \psi_n(x) dx$ are the corresponding transition dipole matrix elements.
For symmetric trapping potentials $W(x)$ the dipole matrix elements $d_n=0$ if $n$ is odd, therefore 
we have chosen the target state $g(x)$ to be an 
antisymmetric function with respect to $x_0$. For an infinite well potential $\epsilon_n=\frac{\hbar^2\pi^2 n^2}{2 m L^2}$,
and the matrix elements can be also calculated analytically for even values of $n$ : $d_n=-4L(\cos(\pi n)n +n)/(\pi(1-2n^2+n^4))$.
The expansion coefficients in the limit of small width of the target state $\alpha \ll L$, asymptotically approach 
$a_n\approx \sqrt{\pi}\alpha^3\exp(-\alpha^2(1+n)^2/4)(1+\exp(\alpha^2 n)(n-1)+n)/4$.

In the present calculations we have chosen $D_0=10^{-3} E_0$.
The initial  states $u_n(x,0), v_n(x,0)$ were not calculated self-consistently, 
 instead we used the approximate  analytical solution Eq.(\ref{anayt}). 
The first six terms in Eq.(\ref{field}) are used as a
 starting approximation for the control field. 
We assumed $\alpha=0.1L$ and set the control time interval 
$T=50 \pi T_1$, where $T_1=\frac{2\pi\hbar}{\epsilon_1}=\frac{4m L^2}{\pi\hbar}$ 
is the time period 
for a particle occupying the ground state of  an infinite well potential. 
For this relatively long control interval the amplitude a quasiparticle, which
occupies the ground state of the potential 
will make about $150$ oscillations.  Therefore, the Rotating Wave Approximation,
used in the derivation of Eq.(\ref{field}) in \cite{squeezing} is 
justified. The optimal control will result in relatively slow Rabi oscillations   of the occupation numbers
$|a_n(t)|^2 \approx |a_n|^2 \sin^2(\frac{\pi t}{2T})$, $n>1$ \cite{squeezing}, where
$a_n(t)=\int u_1(x,t)\psi_n(x)dx$. Similar dynamics can be observed for the quasi-holes 
$b_n(t)=\int v_1(x,t)\psi_n(x)dx$.

The amplitudes $V_n$ and the frequencies $\omega_n$ of the control field were used as an initial input for 
a black box optimization
 using  simulated annealing algorithm. 
As the fitness function for the optimization we used the averaged 
pair potential Eq.(\ref{delta}) on the spatial interval $[L/2-\epsilon_0,L/2+\epsilon_0]$, with $\epsilon_0=0.1 L$,
and $t_0=0.1 T$. 
The simulated annealing optimization resulted in additional $\approx15 \%$ improvement of the averaged magnitude of the
 pair potential $<\Delta>$ with respect to the enhancement, obtained using the non-optimized  analytical solution
 Eq.(\ref{field}).
To reduce the amount of the meshing points  we set  the renormalized chemical potential $\tilde\mu$ to zero, 
therefore limiting the 
amount of the nodes in the initial quasiparticle amplitudes.
We considered the evolution of 
maximum $6$ (3 quasi-electron, and 3 quasi-hole) quasi particles.   
Such a small number was chosen because the 
characteristic single particle frequencies
in a square well potential grow quickly (as $n^2$) that makes numerical integration of TDBdG equations extremely slow. 
The resulting nonequilibrium dynamics of the pair potential $|\Delta(x,t)|$ is shown
 in Fig.\ref{fig1}.  Note we assumed zero temperature.
In Fig.1(a) one can see the gradual spatial localization of the
pair potential near the center of the infinite well potential $x_0=L/2$,
  which occurs at the end of the control interval.
  The local enhancement of the pair potential and the  transition temperature $T_c (x_0,T) \propto \Delta (x_0,T)$  is 
about $300\%$ compared to the initial state.
 In Fig.\ref{fig1}(a) one can see that the enhancement 
is achieved at the price of  decrease of the
$|\Delta(x,t)|$ outside the target area, closer to the walls of the trapping potential. 
In Fig.1 (b,c) we have shown similar simulations for optimal control of 
four and six quasiparticles, and the resulting dynamics of the pair potential. 
As in the case of just two quasiparticles, the pair potential shows a strong enhancement near the center of the well. 
Note, the control field $V(x,t)$ also drives other quasiparticle amplitudes $u_n(x,T)$, $v_n(x,T)$ with $n>1$, but
this does not change much the localization picture.  
One has to note that the enhancement is relatively weaker in the latter two cases shown in Fig.1 (b,c).
The approximate analytical solution for the amplitudes $a_n(t)$ and $b_n(t)$ 
has the period of $4T$, therefore, for the controlled  system $|a_n(t)|^2$ and $|b_n(t)|^2$ will return to their 
initial states at $t=2T$. In real systems
this return will never be perfect, and it will correspond to the loss of 
coherence in the driven system and the energy dissipation. For 
strongly interacting quasiparticles the driven dynamics can significantly 
depart from the approximate analytical solution.
       \begin{figure}
\includegraphics[width=7.cm]{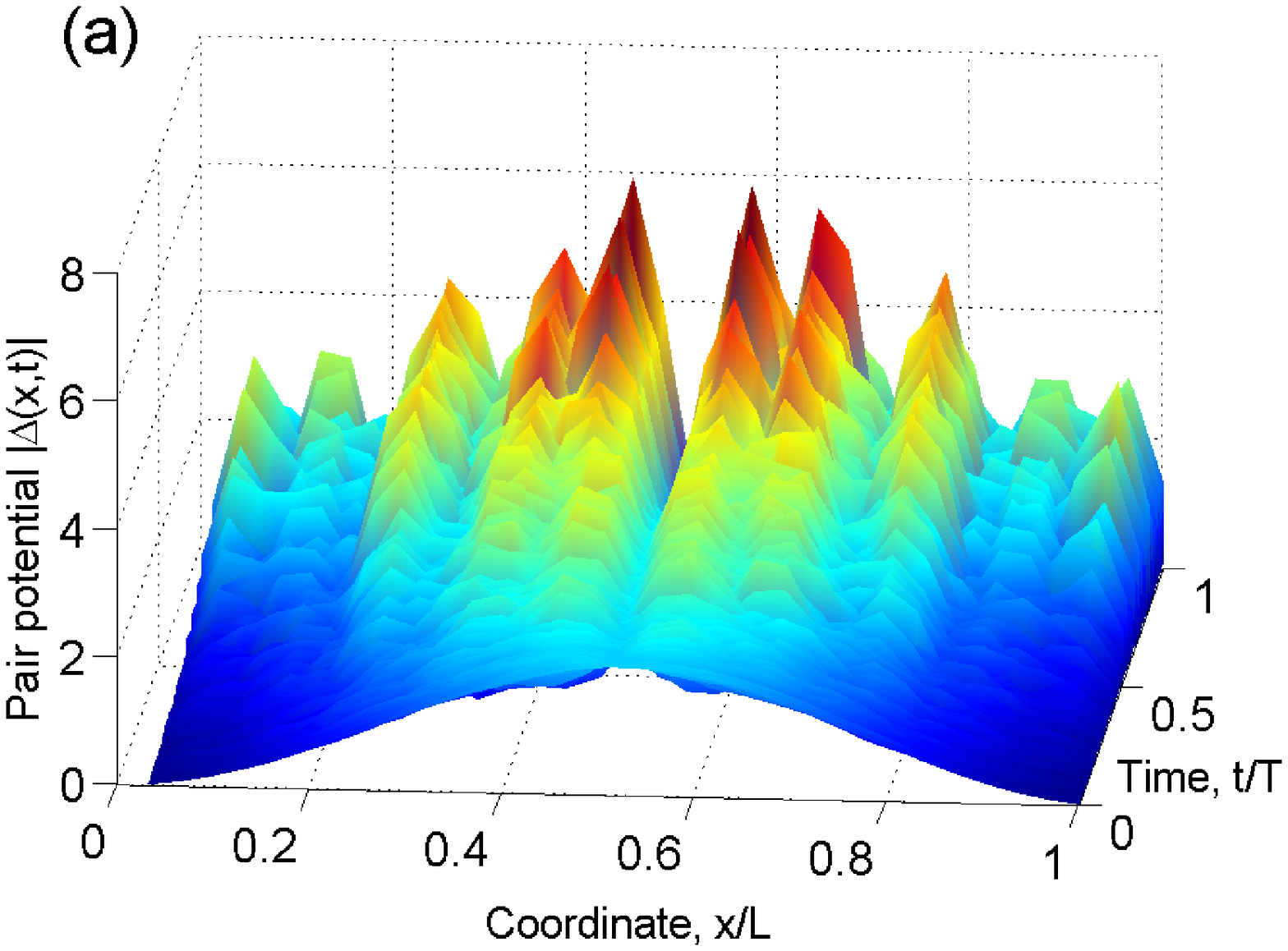}
\includegraphics[width=7.cm]{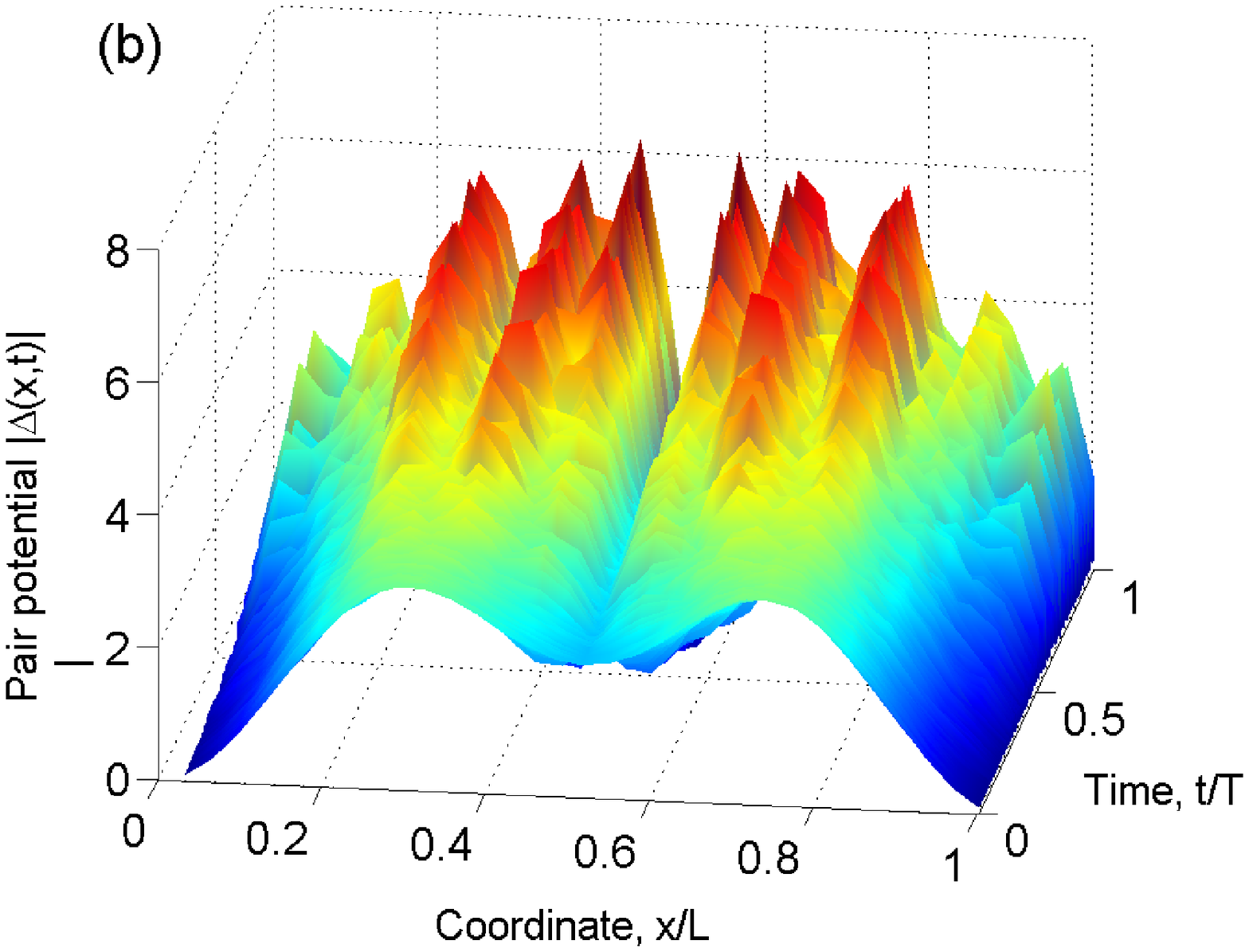}
\includegraphics[width=7.cm]{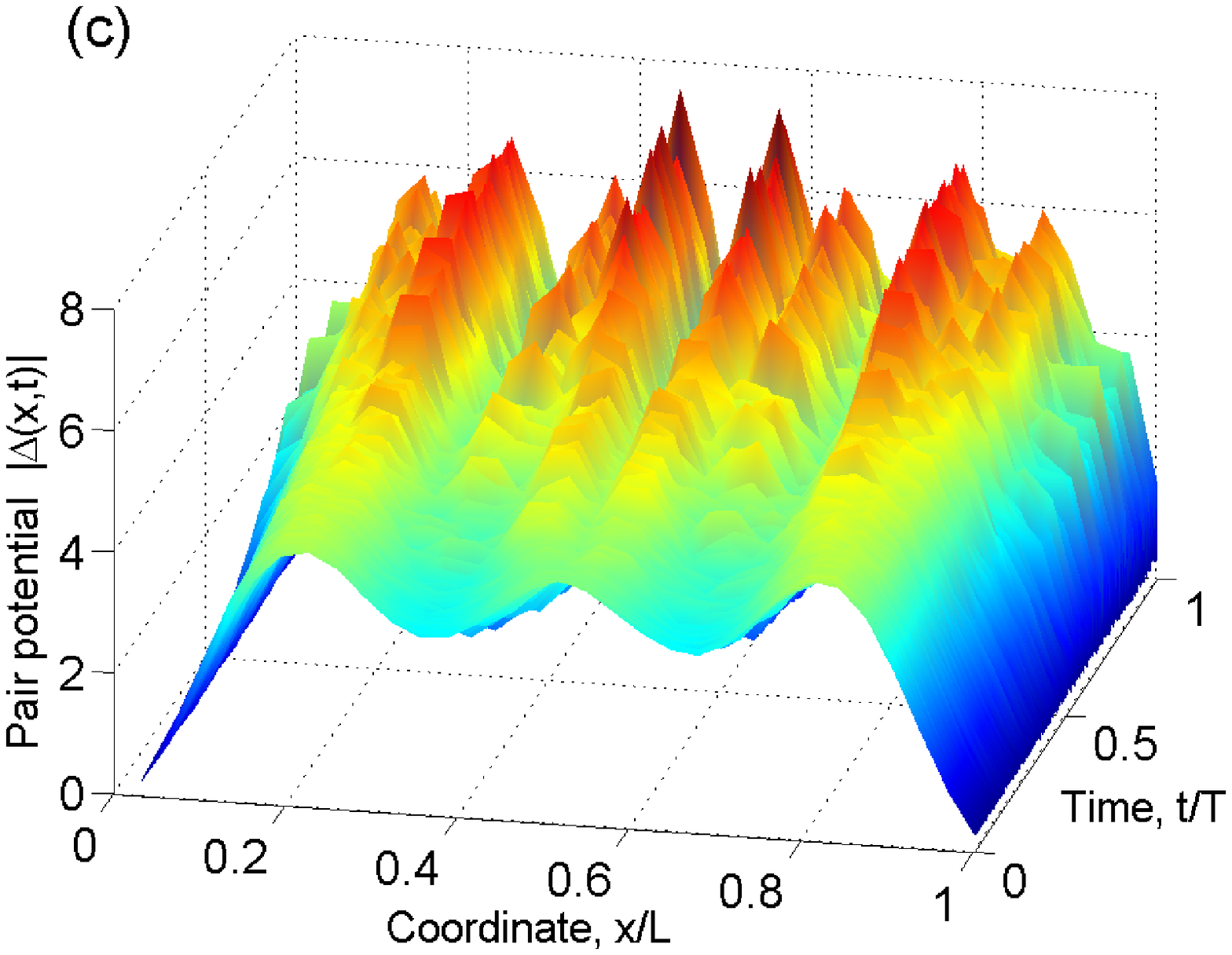}
\caption{\label{fig1} Spatial localization and enhancement of the pair potential $|\Delta(x,t)|$ (arbitrary units) using 
an optimized external control field.  
The optimization was performed with the target of maximizing the fitness function in Eq.(\ref{delta}). 
(a) two quasi-particles in the infinite well  potential (described by the amplitudes $u_1(x,t)$ and $v_1(x,t)$)
(b) four quasi-particles, (c) six quasiparticles.
}
\end{figure}
In superconducting materials 
the pair potential $\Delta$ is relatively small compared to the Fermi energy, and can be
treated as a perturbation. However, this is not always the case, for example, in cold atomic Fermi gases.
Relatively strong coupling between functions $u_n$ and $v_n$ in the presence of
 oscillatory field may lead to chaotic dynamics. In this case
numerically stable control techniques of chaos are necessary
\cite{chaos}.


The proposed mechanism may be applied not only to cold atomic Fermi gases,
but also in solid state quasi-one dimensional superconducting systems.
 However, the thickness of the system should not be too
small, since  the gap enhancement can be suppressed
by the phase slip phenomena  \cite{kopnin}, which is not included in our mean field description.
One may consider an experiment with  quasi one dimensional superconductors, such as a nanowire or a bundle of 
doped carbon nanotubes \cite{nanowire}.   
One may design an optimal control experiment, when the quantity of interest will be 
non-equilibrium conductivity of the bundle at a temperature close to $T_c$. Making analogy 
with the Eliashberg effect \cite{gulian}, one may expect that 
a nonequilibrium perturbation  of the system may increase the effective  critical temperature
(as well as the instantaneous pairing
potential).     
Using an optimal 
control field one may drive periodically the quasiparticle density to be maximal along the bundle near its center, that will 
result in a considerable drop of the resistance of the bundle at times of the spatial localization.
In order to achieve this, the control field should have its polarization, which is transverse to the bundle.

The control field  Eq.(\ref{field}) can be used in the experiment as an initial guess. Then one may use
 the measured nonequilibrium conductivity of the bundle at the target moment $T$ as a fitness function, 
in a similar fashion we used  Eq.(\ref{delta}) for the simulated annealing optimization. 
Let us consider a bundle of boron or alkali-atoms doped carbon nanotubes of width $L=500$nm  \cite{nanowire}. 
We assume the level of doping such that the carriers are having the wavelength of the order of $4$ nm. 
As the initial guess for the control field we may choose 
a field that drives the initial quasiparticle 
wavefunction with the lowest energy $u_n(x)=u_{250}\propto\sin(250\pi x/L)$ to the target 
$u_{250}(x,T)\propto u_n(x,0) g(x)$, where the function
$g(x)$ is defined above. Note the control will be performed rather over the envelope of the quasiparticle wavefunctions.
At the time $T$ the target wavefunction can be represented as 
$u_{250}(x,T)=\sum_{j=1,3,..}\sum_{n=250\pm j}a_{n}u_{n}(x,0)$.
The expansion coefficients of the target wavepacket  $a_n$ 
can be used in the solution Eq.(\ref{field}). 
The same control is applicable for the quasi-hole wavefunction $v_{250}(x)$. Even for a relatively modest density
localization with $\alpha=0.3 L$ for just one pair of quasiparticles with the lowest energy, 
it will correspond to $\approx 300\%$ 
increase of the  term $v^*_{250}u_{250}$ and the overall local increase of the pairing potential by $\approx 1\%$. 
 One can use the
adaptive optimization approach  to further improve the localization of  quasiparticles \cite{teaching}.
There may be some possible obstacles for the proposed experiment. First, relatively strong control field
may result in strong energy dissipation. At the same time 
the control field can not be too weak, since it will increase the duration of the
optimal control interval $T$. The duration of the control interval should not be too long, because 
the loss of coherence, for example, due to collisions of cold atoms, may  reduce the efficiency of the proposed scheme \cite{decoherence_control}.  

To conclude, we presented a new approach to control the local enhancement
of the pair potential.
The enhancement is achieved through  a nonequilibrium spatial localization
of the quasiparticle density in the system.
This mechanism contrasts with  the spatially homogeneous Eliashberg mechanism, which usually
employs a field with only a single frequency. Our method is based on coherent control, which
requires a multi-frequency field with independently tuned field amplitudes.
The suggested control scheme will be effective on the time scale,  shorter than  
the effective decoherence times, which  for Fermi gases can be of the order of milliseconds.
The time-dependent Eq.(\ref{tdbdg}) is
valid for weakly nonequilibrium conditions, since $\Delta({\bf r},t)$ is defined
by using the equilibrium values of the Fermi distribution $f_n$.
Most significant, the electromagnetic field should not have components with
the frequencies over the double gap size, $\hbar\omega_{max}>2|\Delta|$,  because high energy photons 
 can easily break  Cooper pairs. Assuming the equilibrium pairing potential 
$\Delta\approx0.0015$eV in the above described example with the bundle of doped carbon nanotubes, 
one may estimate that the transition frequencies between 
the nearest levels at the Fermi surface (assuming the infinite well potential model)
will be about $\Delta E\approx 0.0007$eV. Therefore operating with a radio frequency field with 
$\hbar\omega_{max}<3\times10^{-3}eV$ one still will be able to create a linear superposition of  states 
$u_{250}(x,T)=a_{251}u_{251}(x,0)+a_{253}u_{253}(x,0)+a_{255}u_{255}(x,0)$, which has a considerable
 ``bump'' near the center of the bundle. We would like to emphasize that direct
microscopic simulations using Eq.(\ref{tdbdg}) will need at least a thousand 
spatial discretization points that makes the problem currently computationally intractable.

\end{document}